\documentclass[english,twocolumn]{revtex4-1}
\usepackage[T1]{fontenc}
\usepackage[latin9]{inputenc}
\usepackage{amsmath}
\usepackage{graphicx}
\usepackage{babel}
\begin{document}
\title{Steady state thermodynamics of ideal gas in shear flow}
\author{Karol Makuch}
\email{kmakuch@ichf.edu.pl}

\affiliation{Institute of Physical Chemistry, Polish Academy of Sciences Kasprzaka
44/52, 01-224 Warszawa}
\author{Konrad Gi\.{z}y\'{n}ski}
\affiliation{Institute of Physical Chemistry, Polish Academy of Sciences Kasprzaka
44/52, 01-224 Warszawa}
\author{Anna Macio\l ek}
\affiliation{Institute of Physical Chemistry, Polish Academy of Sciences Kasprzaka
44/52, 01-224 Warszawa}
\affiliation{Max-Planck-Institut f{\"u}r Intelligente Systeme Stuttgart, Heisenbergstr.~3,
D-70569 Stuttgart, Germany}
\author{Pawe\l{} J. \.{Z}uk}
\affiliation{Institute of Physical Chemistry, Polish Academy of Sciences Kasprzaka
44/52, 01-224 Warszawa}
\affiliation{Department of Physics, Lancaster University, Lancaster LA1 4YB, United
Kingdom}
\author{Robert Ho\l yst}
\email{rholyst@ichf.edu.pl}

\affiliation{Institute of Physical Chemistry, Polish Academy of Sciences Kasprzaka
44/52, 01-224 Warszawa}
\begin{abstract}
Equilibrium thermodynamics describes the energy exchange of a body
with its environment. Here, we describe the global energy exchange
of an ideal gas in the Coutte flow in a thermodynamic-like manner.
We derive a fundamental relation between internal energy as a function
of parameters of state. We analyze a non-equilibrium transition in
the system and postulate the extremum principle, which determines
stable stationary states in the system. The steady-state thermodynamic
framework resembles equilibrium thermodynamics. 
\end{abstract}
\maketitle

\section{Introduction}

There is mounting evidence that the exchange of energy of a macroscopic
steady state system with its environment can be described in an equilibrium
thermodynamic-like fashion. It has been considered in the 90s of the
twentieth century for systems with chemical reactions that may constantly
occur in time \citep{ross1988thermodynamics,hunt1990thermodynamic,ross1992thermodynamic}.
Such a steady state situation goes beyond equilibrium thermodynamics
which excludes any macroscopic flows and currents \citep{Thermodynamics_and_an_Introduction_to_Thermostatistics_2ed_H_Callen}.
Around a similar time, the thermodynamic-like approach was introduced
for systems with shear-flow \citep{evans1980thermodynamics,hanley1982thermodynamics,evans1983computer,evans1986rheology}
and is still under development \citep{daivis2003steady,taniguchi2004steady,daivis2008thermodynamic,baranyai2018thermodynamic}.
Another class of macroscopic systems for which the attempt to introduce
thermodynamic description has been undertaken is systems with heat
flow \citep{nakagawa2019global,nakagawa2022unique,Zhang2021continuous,holyst2022thermodynamics,Holyst2023Steady}.

The above approaches to formulate steady state thermodynamics focus
on chemical reactions and shear flow, both in homogeneous temperature
profiles or systems with heat flow without chemical reactions and
macroscopic flows. If steady state thermodynamics is ever formulated
on general grounds, it is now at its inception. About two decades
ago, Oono and Paniconi \citep{oono1998steady} and Sasa and Hatano
\citep{sasa2006steady} postulated a thermodynamic-like description
based on a general footing. However, because the descriptions were
postulated, they require validation and further investigation on the
eventual limitation. In particular, it is not clear whether the nonequilibrium
entropy defined by the integral of the local entropy density over
the volume of the system \citep{nakagawa2022unique} or rather the
excess heat-based entropy \citep{oono1998steady} should be used to
construct the principles of steady-state thermodynamics. These two
entropies are not equivalent \citep{holyst2022thermodynamics}. It
shows that the steady state thermodynamics is far from complete, and
the core fundamental questions still remain open \citep{Holyst2023Steady,nakagawa2022unique}.
It is worth mentioning stochastic thermodynamics, which focuses on
thermodynamic notions for the system on the level of individual trajectories
\citep{seifert2012stochastic,ding2022unified}. Here we focus on a
reduced, macroscopic description of a steady state system \citep{speck2021modeling}.

Allowing various macroscopic constant fluxes drives the system from
equilibrium to a steady state and opens up phenomena that eludes equilibrium
thermodynamics. Take for example a quiescent liquid in a uniform gravitational
field in equilibrium. This situation assumes a uniform temperature
and no macroscopic flows, which, together with the equations of state,
determine the thermodynamic parameters at each point of the system.
Allowing a vertical or horizontal heat flow radically changes the
situation. The flow of heat combined with the gravitational force
can cause regular mass movement, either because the denser liquid
is at the top and the thinner is at the bottom or because gravity
unevenly acts on areas with different horizontal densities. This unstable
situation causes a mass movement in the atmosphere, oceans, planets,
and stars \citep{bodenschatz2000recent}. The core feature of this
phenomenon is the coupling between the heat flow and the mass flow.
However, the question arises whether some reduced description for
non-equilibrium steady states also emerges from hydrodynamics. Can
it take the shape of equilibrium thermodynamics, in which the system's
behavior is described by only a few parameters and the rules of extremum
containing only these parameters?

Here we present a thermodynamic-like description of a system with
heat and mass flow coupling. We consider ideal gas in shear flow with
a dissipative temperature profile shown schematically in Fig. \ref{fig:Ideal-gas-in}.
We introduce the first law for this system that describes different
ways of exchanging the system's energy with its environment. We also
consider a movable wall as a thermodynamic constraint in the system.
We introduce the extremum principle that determines the position of
the wall. We show that there is a critical shear in the system above
which the equilibrium position of the internal wall becomes unstable,
and the system undergoes a nonequilibrium second order phase transition.
We give a complete thermodynamic-like description of this steady state
system in which the position of the movable wall is a thermodynamic
constraint. For the vanishing shear flow, the steady state extremum
principle directly reduces to the minimum principle in the corresponding
problem in thermodynamics.

\section{System}

\begin{figure}
\includegraphics[width=8.5cm]{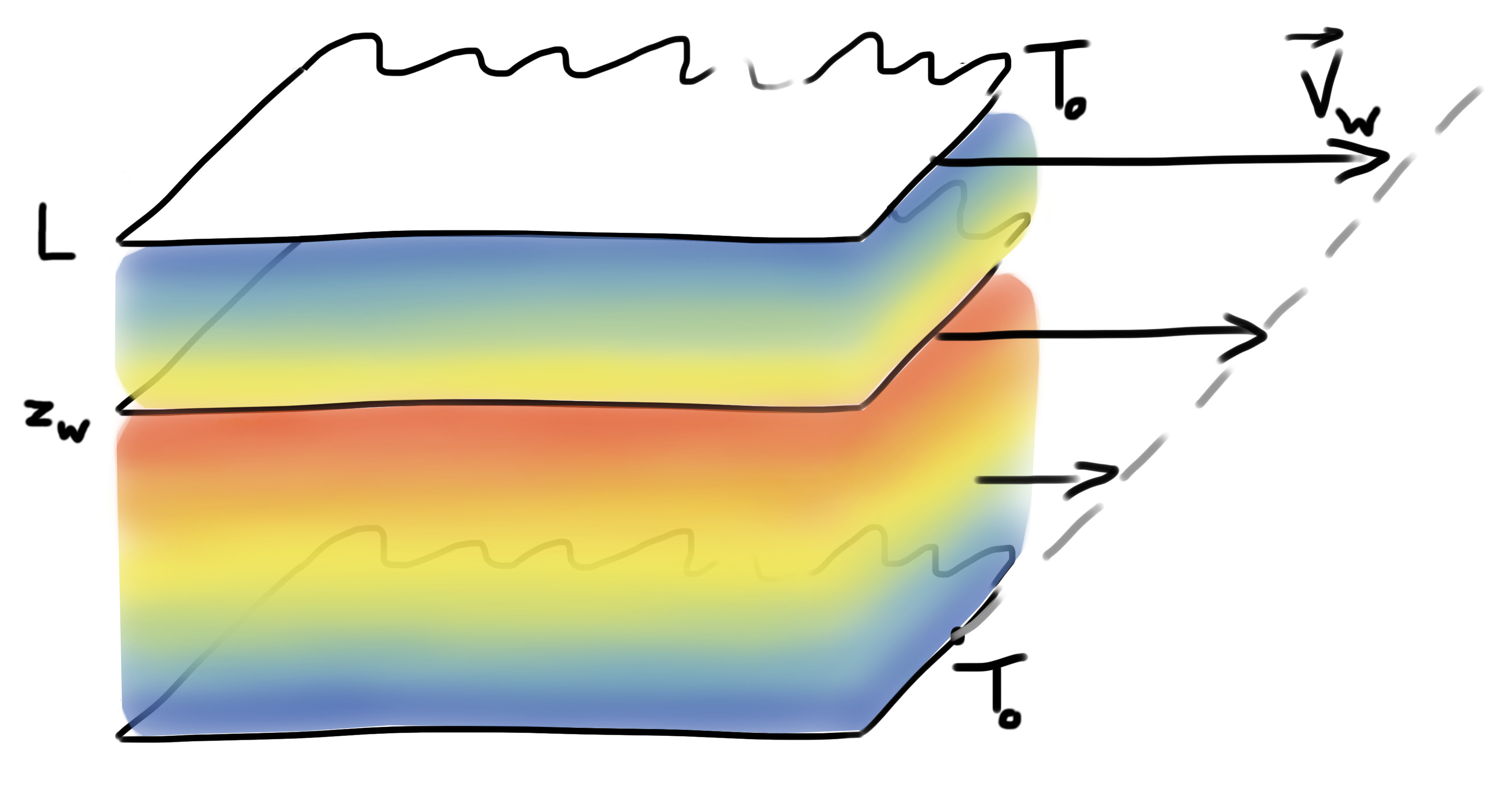}\caption{\label{fig:Ideal-gas-in}Ideal gas in shear flow. The gas is confined
by the immobile bottom wall and by the upper wall moving with velocity
$V_{w}$. There is also a thermally insulating and moving internal
wall that divides the gas into two subsystems. Shearing motion causes
energy dissipation in the system, which changes the temperature profile.
The heat flows out from the system through the confining walls at
temperature $T$. Colors schematically show the temperature profile
(red - hotter, blue - colder).}
\end{figure}

We consider ideal monoatomic gas between two two parallel walls at
walls' temperature $T_{0}$, as shown in Fig. \ref{fig:Ideal-gas-in}.
The upper wall moves with speed $V_{w}$ moving in $x$-direction.
The system is described by irreversible thermodynamic equations: the
continuity equation, Navier-Stokes equation, energy balance and two
equations of state \citep{Groot_Mazur_Non-equilibrium_thermodynamics}.
We assume that the system is translationally invariant in $x$ and
$y$ directions, therefore the fields depend only on $z$ coordinate.
In particular, equations of state are given by,

\begin{equation}
p\left(z\right)=n\left(z\right)k_{B}T\left(z\right),\label{eq:pressur eos}
\end{equation}
\begin{equation}
u\left(z\right)=\frac{3}{2}n\left(z\right)k_{B}T\left(z\right),\label{eq:temperature eos}
\end{equation}
with pressure $p\left(z\right),$ volumetric particle number density
$n\left(z\right)$, temperature $T\left(z\right)$, Boltzmann constant
$k_{B}$, and the volumetric energy density, $u\left(z\right)$. The
velocity field also depends on $z$ coordinate only and is oriented
in the direction determined by the moving wall, 
\begin{equation}
\mathbf{v}\left(x,y,z\right)={\bf e}_{x}v\left(z\right).\label{eq:velocity symmetry}
\end{equation}

\section{Steady state}

We describe the system within linear response theory \citep{Groot_Mazur_Non-equilibrium_thermodynamics}.
Therefore, the system is locally described by equation of states (\ref{eq:pressur eos})
and (\ref{eq:temperature eos}) supplemented with mass, momentum and
energy conservation. We neglect bulk viscosity. We assume that the
shear viscosity coefficient $\eta$ does not depend on the local state
of the gas, so the viscosity does not depend on local density and
temperature. The translational symmetry for the pressure, $p\left({\bf r}\right)=p\left(z\right)$,
and for the velocity given by Eq. (\ref{eq:velocity symmetry}) reduces
the $z$-component of the Navier-Stokes equation, 
\begin{equation}
\rho\left(\partial_{t}{\bf v}+\left({\bf v}\cdot\nabla\right){\bf v}\right)=-\nabla p+\eta\Delta{\bf v}+\frac{1}{3}\eta\nabla\left(\nabla\cdot{\bf v}\right),\label{eq:NavierStokes}
\end{equation}
to the formula, $dp\left(z\right)/dz=0$, while the $x$-component
of the Navier-Stokes equation gives $0=\eta d^{2}v\left(z\right)/dz^{2}$.
Therefore the pressure in the system is homogeneous,
\begin{equation}
p\left(z\right)=p,\label{eq:homoegeinty p}
\end{equation}
and, using the boundary conditions $v\left(0\right)=0$ and $v\left(L\right)=V_{w}$,
there is a shear flow in the system, 
\begin{align}
v\left(z\right) & =\frac{z}{L}V_{w}.\label{eq:v shear flow}
\end{align}

The continuity equation, 
\begin{equation}
\partial_{t}\rho+{\bf v}\cdot\nabla\rho=-\rho\nabla\cdot{\bf v},\label{eq:continuity equation}
\end{equation}
with mass density $\rho\left(z\right)$ and the above velocity field
is satisfied automatically for any density profile. The energy balance
equation is given by, 
\begin{align}
\rho\left(\partial_{t}u_{m}+{\bf v}\cdot\nabla u_{m}\right) & =\kappa\Delta T-p\nabla\cdot{\bf v}\nonumber \\
 & +2\eta\mathring{\left[\text{\ensuremath{\nabla}}{\bf v}\right]}^{s}:\mathring{\left[\text{\ensuremath{\nabla}}{\bf v}\right]}^{s},\label{eq:internal energy}
\end{align}
where $\mathring{\left[\text{\ensuremath{\nabla}}{\bf v}\right]}_{ij}^{s}=\left(\nabla_{i}\mathbf{v}_{j}+\nabla_{j}\mathbf{v}_{i}-\delta_{ij}\frac{2}{3}\nabla\cdot{\bf v}\right)/2$
denotes the symmetrized and traceless velocity gradient matrix, the
colon is the double contraction, $\delta_{ij}$ denotes the Kronecker
delta, and $u_{m}=u/\rho$ is the mass density of the internal energy.
Utilizing the shear velocity profile (\ref{eq:v shear flow}) and
the translational invariance of the internal energy density, $u_{m}\left(z\right)$,
the energy balance equation reduces to the equation for the temperature
profile

\begin{equation}
0=\kappa\frac{d^{2}}{dz^{2}}T\left(z\right)+\eta\left(\frac{V_{w}}{L}\right)^{2}.\label{eq:red en balance eq}
\end{equation}
The above equation has the following form, $0=\kappa\frac{d^{2}}{dz^{2}}T\left(z\right)+\lambda$,
where $\lambda$ is the volumetric heating rate,
\begin{equation}
\lambda=\eta\left(\frac{V_{w}}{L}\right)^{2}.\label{eq:def lambda}
\end{equation}
We also introduce the dimensionless parameter
\[
D\equiv\frac{\eta V_{w}^{2}}{\kappa T_{0}}.
\]

\section{Shear flow without the internal wall}

We first consider the shear flowing system shown in Fig. \ref{fig:Ideal-gas-in},
but without the internal wall. In this situation, the boundary conditions
are that on the upper and lower surface and the temperature profile
is given by,

\begin{equation}
T\left(z\right)=T_{0}\left[-\frac{1}{2}\frac{\lambda L^{2}}{\kappa T_{0}}\left(\frac{z}{L}\right)^{2}+\frac{1}{2}\frac{\lambda L^{2}}{\kappa T_{0}}\frac{z}{L}+1\right].\label{eq:temp profile}
\end{equation}
Number of particles in the surface area $A$ of the translationally
invariant system is determined by the particle density, $N=A\int_{0}^{L}dz\,n\left(z\right)$,
which with the mechanical equation of state (\ref{eq:pressur eos})
and the above temperature profile gives
\begin{equation}
p=\frac{Nk_{B}T_{0}}{AL}\psi\left(\frac{\eta V_{w}^{2}}{\kappa T_{0}}\right),\label{eq:pressure}
\end{equation}
where
\begin{equation}
\psi(D)\equiv\frac{D}{8\sqrt{\frac{D}{8+D}}\tanh^{-1}\sqrt{\frac{D}{8+D}}}.\label{eq:func}
\end{equation}
Combination of equations of state (\ref{eq:pressur eos}) and (\ref{eq:temperature eos})
gives the following internal energy density $u\left(z\right)=3p\left(z\right)/2.$
Total internal energy in the system is given by 
\begin{equation}
U=A\int_{0}^{L}dz\,u\left(z\right)=\frac{3}{2}ALp,\label{eq:energy and pressure}
\end{equation}
where we used the fact, that the pressure in the system is constant.
Utilizing Eq. (\ref{eq:pressure}) in the above equation we get,
\begin{equation}
U=\frac{3}{2}Nk_{B}T_{0}\psi\left(\frac{\eta V_{w}^{2}}{\kappa T_{0}}\right).\label{eq:U by psi}
\end{equation}
We determine the kinetic energy due to macroscopic motion, $E_{k}=A\int_{0}^{L}dz\,mn\left(z\right)v^{2}\left(z\right)/2,$
where $m$ is a mass of a single ideal gas molecule. We obtain
\begin{equation}
E_{k}=mN\frac{V_{w}^{2}}{2}\varepsilon\left(\frac{\eta V_{w}^{2}}{\kappa T_{0}}\right),\label{eq:ek by eps}
\end{equation}
where 
\begin{equation}
\varepsilon\left(D\right)=\frac{2}{D}\left(\frac{4+D}{4}\frac{1}{\psi\left(D\right)}-1\right).\label{eq:def eps}
\end{equation}

Utilizing Eqs. (\ref{eq:U by psi}), (\ref{eq:ek by eps}) and (\ref{eq:def eps})
we obtain the relation between the kinetic and internal energy,
\begin{equation}
E_{k}\frac{\eta}{mN\kappa T_{0}}=\frac{4+\frac{\eta V_{w}^{2}}{\kappa T_{0}}}{4}\frac{\frac{3}{2}Nk_{B}T_{0}}{U}-1.\label{eq:ek and u}
\end{equation}

\section{Transition between steady states and energy balance}

We consider transitions between steady states in the system. The system
is in one steady state at time $t_{i}$ , after which we slightly
change one or more control parameters that appear in (or are related
to) the governing equations. For example, we slowly modify the velocity
of the wall, the position of the upper wall (volume $V$) or the external
temperature $T_{0}$. In particular, we increase the velocity of the
wall by $V_{w}\left(t\right)=V_{w}+\left(t-t_{i}\right)dV_{w}/\left(t_{f}-t_{i}\right)$
for $t_{i}<t<t_{f}$, which modifies velocity of the wall from $V_{w}$
to $V_{w}+dV_{w}$. This induces a time dependent hydrodynamic field.
After time $t_{f}$ the system reaches another steady state close
to the previous one due to a small change in the control parameters.
We focus on steady, nonequilibrium states with a constant number of
particles in the system.

Previously we have analyzed the energy balance for an ideal gas in
a heat flow \citep{holyst2022thermodynamics} during the transition
between steady states. The considerations necessary for Couette flow
go along the same line. We do not repeat them but give a sketch. We
monitor the change of the energy between times $t_{i}$ and $t_{f}$,
\[
dU=U\left(t_{f}\right)-U\left(t_{i}\right)=\int_{t_{i}}^{t_{f}}dt\,\frac{dU\left(t\right)}{dt},
\]
where
\begin{equation}
U\left(t\right)=\int_{V(t)}d^{3}r\,\rho\left({\bf r},t\right)u\left({\bf r},t\right),\label{eq:internal energy-1}
\end{equation}
and $\rho\left({\bf r},t\right)$ is the density and $u\left({\bf r},t\right)$
is the internal energy density per unit mass at position ${\bf r}$
and time $t$. We then represent the time derivative of the above
integral utilizing the internal energy balance equation (\ref{eq:internal energy}).
Applying analysis from \citep{holyst2022thermodynamics} leads to
the conclusion that the system exchanges the energy with the environment
according to the following formula,
\begin{equation}
dU=\mkern3mu\mathchar'26\mkern-12mu dQ+\mkern3mu\mathchar'26\mkern-12mu dW+\mkern3mu\mathchar'26\mkern-12mu dD,\label{eq:first law int en}
\end{equation}
with the excess heat defined by 
\begin{equation}
\mkern3mu\mathchar'26\mkern-12mu dQ=-\int_{t_{i}}^{t_{f}}dt\int_{\partial V\left(t\right)}d^{2}r\,\hat{n}\cdot\left(J_{q}\left(t\right)-J_{q,st}\left(t\right)\right),\label{eq:excess heat}
\end{equation}
where $\hat{n}$ is the vector normal to the surface of volume $V$
and $J_{q}\left(t\right)$ is the heat flux in the system, with the
work related to the expansion of the gas defined by
\begin{equation}
\mkern3mu\mathchar'26\mkern-12mu dW=-\int_{t_{i}}^{t_{f}}dt\int_{V\left(t\right)}d^{3}r\,p\nabla\cdot{\bf v},\label{eq:volumetric work}
\end{equation}
and the excess dissipation defined by,
\begin{equation}
\mkern3mu\mathchar'26\mkern-12mu dD=\int_{t_{i}}^{t_{f}}dt\int_{V\left(t\right)}d^{3}r\,\left(\Pi:\left[\nabla{\bf v}\right]-\Pi_{st}\left(t\right):\left[\nabla{\bf v}_{st}\left(t\right)\right]\right),\label{eq:excess dissipation}
\end{equation}
where $\left[\nabla{\bf v}\right]_{ij}=\nabla_{i}{\bf v}_{j}$ is
the gradient velocity matrix. The subscript $st$ in the above equations
means that the quantity should be taken at a stationary state that
corresponds to the control parameters at time $t$. The first term
in the internal energy balance (\ref{eq:first law int en}), $\mkern3mu\mathchar'26\mkern-12mu dQ$,
is the heat on top of the steady state heat exchanged during the transition
between steady states \citep{oono1998steady}. The second term in
(\ref{eq:first law int en}) is the volumetric work performed by that
wall (more specifically, by the perpendicular component of the force
of the wall) when the system is compressed. The third term $\mkern3mu\mathchar'26\mkern-12mu dD$
is the energy dissipation in the system on top of the steady state
dissipation.

Because the pressure in a steady state is spatially uniform, the work
from Eq. (\ref{eq:volumetric work}) for slow transitions between
steady states reduces as follows, $\mkern3mu\mathchar'26\mkern-12mu dW\approx-p\int_{t_{i}}^{t_{f}}dt\int_{V\left(t\right)}d^{3}r\,\nabla\cdot{\bf v}$.
The integrals of the divergence with the Gauss theorem reduce to the
change of the volume, and the work reduces to the volumetric work
that also appears in equilibrium thermodynamics,

\[
\mkern3mu\mathchar'26\mkern-12mu dW=-pdV.
\]

To determine the exchange of the kinetic energy, we proceed along
the same line as for the internal energy above. Thus we use the kinetic
energy balance equation \citep{Groot_Mazur_Non-equilibrium_thermodynamics},
\begin{equation}
\partial_{t}\frac{1}{2}\rho{\bf v}^{2}=-\nabla\cdot\left(\frac{1}{2}\rho{\bf v}^{2}{\bf v}+P\cdot{\bf v}\right)+P:\left[\nabla{\bf v}\right],\label{eq:kin energy balance}
\end{equation}
where the pressure tensor $P$ in the system is given by
\begin{align*}
P_{ij} & =p\delta_{ij}+\Pi_{ij},\\
\Pi & =-2\eta\mathring{\left[\text{\ensuremath{\nabla}}{\bf v}\right]}^{s}.
\end{align*}
Analysis of the change of the kinetic energy $dE_{k}=E_{k}\left(t_{f}\right)-E_{k}\left(t_{i}\right)=\int_{t_{i}}^{t_{f}}dt\,\frac{dE_{k}\left(t\right)}{dt}$
leads to the conclusion that during the transition between steady
states, the kinetic energy changes according to the following equation,
\begin{equation}
dE_{k}=\mkern3mu\mathchar'26\mkern-12mu dW_{w}-\mkern3mu\mathchar'26\mkern-12mu dD,\label{eq:first law kin en}
\end{equation}
where 
\[
\mkern3mu\mathchar'26\mkern-12mu dW_{w}=\int_{t_{i}}^{t_{f}}dt\int_{\partial V\left(t\right)}d^{2}r\,\hat{n}\cdot\left(\Pi\cdot{\bf v}-\Pi_{st}\left(t\right)\cdot{\bf v}_{st}\left(t\right)\right)
\]
According to the above equations, the kinetic energy may change in
two ways. $\mkern3mu\mathchar'26\mkern-12mu dW_{w}$ is the work performed
by the upper wall on top of the steady work performed to keep the
shearing steady state. It may be called the excess shear work of the
wall. This effect has been recently discussed by Baranyai \citep{baranyai2018thermodynamic}.

The differential of the total energy, $E=U+E_{k}$ follows from Eqs.
(\ref{eq:first law int en}) and (\ref{eq:first law kin en}) and
is given by,
\begin{equation}
dE=\mkern3mu\mathchar'26\mkern-12mu dQ-pdV+\mkern3mu\mathchar'26\mkern-12mu dW_{w}.\label{eq:first law total en}
\end{equation}
The total energy changes as a result of the excess heat $\mkern3mu\mathchar'26\mkern-12mu dQ$,
volumetric work $-pdV$, and the excess shear work $\mkern3mu\mathchar'26\mkern-12mu dW_{w}$.
Contrary to the energy balance equations (\ref{eq:first law int en})
and (\ref{eq:first law kin en}), there is no excess dissipation $\mkern3mu\mathchar'26\mkern-12mu dD$
defined by Eq. (\ref{eq:excess dissipation}) in the above total energy
balance equation. It follows that the excess dissipation $\mkern3mu\mathchar'26\mkern-12mu dD$
describes an internal transfer of internal and kinetic energy. During
the steady state transition, part of the kinetic energy changes to
internal energy through the kinetic energy dissipation that occurs
on top of the steady state dissipation.

\section{Nonequilibrium entropy}

By introducing 
\begin{equation}
\mkern3mu\mathchar'26\mkern-12mu dP\equiv\mkern3mu\mathchar'26\mkern-12mu dQ+\mkern3mu\mathchar'26\mkern-12mu dD\label{eq:def dP}
\end{equation}
we obtain the internal energy balance (\ref{eq:first law int en})
in the following form, 
\begin{equation}
dU=\mkern3mu\mathchar'26\mkern-12mu dP-pdV.\label{eq:ine t through P and vol work}
\end{equation}
The above formula holds in the space of the control parameters, which
are temperature $T$, the volume of the system $V$, and the speed
of the moving wall $V_{w}$. We do not change the number of particles
$N$. The above formula has the same form as the internal energy of
ideal gas without the macroscopic kinetic energy but with volumetric
heating \citep{holyst2022thermodynamics,Zhang2021continuous}. Because
the pressure in a shear flow system is constant, Eq. (\ref{eq:homoegeinty p}),
and due to the ideal gas mechanical equation of state, Eq. (\ref{eq:energy and pressure}),
we obtain,
\begin{equation}
p=\frac{2}{3}\frac{U}{V}.\label{eq:pressure mapping}
\end{equation}

We notice the following two facts. First, the internal energy balance
in the transition between steady states given by Eq. (\ref{eq:ine t through P and vol work})
is very similar to the equilibrium thermodynamics expression $dU_{eq}=\mkern3mu\mathchar'26\mkern-12mu dQ_{eq}-pdV$.
We see that the differential $\mkern3mu\mathchar'26\mkern-12mu dP$
plays the role of the heat differential in equilibrium thermodynamics.
Second, the expression for pressure (\ref{eq:pressure mapping}) is
also the equilibrium ideal gas pressure mechanical equation, $p=2U_{eq}/3V.$
The above two similarities between steady state and the equilibrium
thermodynamic differentials suggest that there exists the integrating
factor for $\mkern3mu\mathchar'26\mkern-12mu dP$ differential, 
\begin{equation}
\mkern3mu\mathchar'26\mkern-12mu dP=T^{*}dS^{*},\label{eq:int factor for dP}
\end{equation}
where nonequilibrium (effective) entropy is given by the following
fundamental equation for the internal energy,
\begin{equation}
S^{*}\left(U,V\right)=Nk_{B}\log\left[\frac{V}{N}\left(\frac{U}{N}\right)^{3/2}\right]+Ns_{0},\label{eq:int e fundamental relation}
\end{equation}
while the nonequilibrium temperature 
\begin{equation}
T^{*}=\frac{\partial U\left(S^{*},V\right)}{\partial S^{*}}.\label{eq:T star by partial derivative}
\end{equation}
The pressure is determined by $p=-\partial U\left(S^{*},V\right)/\partial V$.
Whether the differential $\mkern3mu\mathchar'26\mkern-12mu dP$ has
integrating factor (is exact) can also be shown directly by proving
that the differential $\mkern3mu\mathchar'26\mkern-12mu dP=dU/T^{*}\left(U,V,V_{w}\right)+p\left(U,V,V_{w}\right)/T^{*}\left(U,V,V_{w}\right)dV+0\cdot dV_{w}$
has equal derivatives, $\partial\left(1/T^{*}\right)/\partial V=\partial\left(p/T^{*}\right)/\partial U$
and that the differential does not explicitly depend on $V_{w}$ \citep{Spivak_M_Calculus_on_manifolds}.

\section{System with internal wall}

Now we focus on the system with the internal wall shown in Fig. \ref{fig:Ideal-gas-in}.~The
wall splits the system into two parts, upper 1 and lower 2. In a steady
state, the velocity field in the system is the shear flow given by
Eq. (\ref{eq:v shear flow}). Therefore the velocity of the internal
wall is $V_{w}^{s}=V_{w}z_{w}/L$, where $z_{w}$ is the position
of this wall. The wall is adiabatically insulating, which changes
one of the boundary conditions. There is no heat flow through the
wall. Without the wall the temperature profile would be given by the
symmetric parabola, Eq. (\ref{eq:temp profile}), but the insertion
of the adiabatic wall in the middle of the system leads to the situation
with half of the parabolic profile. Using this observation, we conclude
that from the perspective of the heat flow equation, we can use the
solution without the internal wall to determine the energy of subsystem
$2$ as follows, $U_{2}\left(T,A,L_{2},V_{w,2}^{u}\right)=U\left(T,A,2L_{2},2V_{w,2}^{u}\right)/2$,
where $V_{w,2}^{u}$ is the speed of the upper wall in subsystem 2
and $L_{2}$ is the vertical length of the subsystem. Because the
change to the moving reference frame does not change the internal
energy, we can deduce the internal energy of the upper system in a
similar manner. The internal energy in both subsystems is given by,
\begin{align*}
 & U_{i}\left(T,A,L_{i},V_{w,i}^{u}-V_{w,i}^{l}\right)\\
 & =\frac{1}{2}U\left(T,A,2L_{i},2\left(V_{w,i}^{u}-V_{w,i}^{l}\right)\right)\text{ for }i=1,2,
\end{align*}
where $V_{w,i}^{u}$ is the speed of the upper wall in subsystem $i$,
and $V_{w,i}^{l}$ is the speed of the lower wall in subsystem $i$.
We have $V_{w,2}^{l}=0$ and $V_{w,2}^{u}=V_{w,1}^{l}$.

Using the properties of the equations of state for ideal gas we can
show that the differentials
\begin{equation}
\mkern3mu\mathchar'26\mkern-12mu dP_{i}\equiv dU_{i}^{int}+p_{i}dV_{i}\text{ for }i=1,2,\label{eq:def dP i through dU dW}
\end{equation}
have integrating factors in agreement with Eq. (\ref{eq:int factor for dP}),
\begin{equation}
\mkern3mu\mathchar'26\mkern-12mu dP_{i}=T_{i}^{*}dS_{i}^{*}\text{ for }i=1,2.\label{eq:integrating factor}
\end{equation}
On the other hand, the analysis of the energy balance similar to the
one presented in the previous section leads to the conclusion that
$\mkern3mu\mathchar'26\mkern-12mu dP_{1/2}=\mkern3mu\mathchar'26\mkern-12mu dQ_{1/2}+\mkern3mu\mathchar'26\mkern-12mu dD_{1/2}$
is the sum of excess heat and excess dissipation in each subsystem.
The differentials $\mkern3mu\mathchar'26\mkern-12mu dQ_{1/2}$ and
$\mkern3mu\mathchar'26\mkern-12mu dD_{1/2}$ are determined by a direct
application of formulas (\ref{eq:excess heat}) and (\ref{eq:excess dissipation})
with the replacement of the volume of integration with the volume
for a particular subsystem.

Until now, we have discussed the energy balance independently for
each subsystem. But it is particularly interesting to consider a perpendicular
motion of the internal wall. As follows from the dynamics, the normal
component of the pressure tensor on the wall in a steady state comes
solely from hydrostatic pressure. The wall may move vertically only
due to the pressure difference, $p_{1}\neq p_{2}.$ It's natural position
is where these pressures are balanced. Therefore the knowledge of
the effective entropy $S^{*}$ and volume for each subsystem is sufficient
to determine the force acting on the wall. The wall's horizontal velocity
enters the problem indirectly. The shear velocity plays the role of
energy dissipation. The speed of the wall appears in the energy balance
equation (\ref{eq:red en balance eq}) in the source term. It thus
plays the role of volumetric heating, $\lambda\equiv\eta\left(V_{w}/L\right)^{2}$.

The ideal gas with volumetric heating and an internal wall has already
been considered previously \citep{Zhang2021continuous}. Such a system
exhibits continuous nonequilibrium phase transition. Similarly, the
system with macroscopic kinetic energy shown in Fig. \ref{fig:Ideal-gas-in}
will exhibit the phase transition as well for the following reason.
Let's focus on the position of the internal wall $z_{w}$ and upper
wall velocity, $V_{w,2}^{u}$, keeping the external walls temperature
$T$, the total volume of the system, $L_{1}+L_{2}=L$, and other
parameters constant. Number of particles in both subsystems is equal,
$N_{1}=N_{2}$. Because the hydrostatic pressure gives the sole force
normal to the walls, the motion of the internal wall can be determined
by the fundamental relations (\ref{eq:int e fundamental relation})
for each subsystem. The pressure follows from $p_{i}=p_{i}\left(U_{i},V_{i},N_{i}\right)$.
Moreover, the internal energy in steady state is determined by Eq.
(\ref{eq:U by psi}). The equivalence is explicit after comparing
Eq. (\ref{eq:U by psi}) to the expression (5) from the reference
\citep{Zhang2021continuous}. This reasoning leads to the conclusion
that from the perspective of the position of the internal wall, the
steady state behavior of both subsystems is the same. The speed of
the upper wall for the system with kinetic energy must be related
to the volumetric heating $\lambda$ from \citep{Zhang2021continuous}
by $\lambda\equiv\eta\left(V_{w}/L\right)^{2}$.

Because there is one-to-one correspondence, we conclude that once
the upper wall's speed increases, the shear flow volumetrically heats
the system. For the value, $\lambda L^{2}/4\kappa T\approx4.55344$,
equivalently $\eta V_{w}^{2}/4\kappa T_{0}\approx4.55344$, the central
position of the wall ceases to be stable. Above the critical speed,
the wall spontaneously chooses one of the two stable positions and
moves away from the center. 

As we show below, it is possible to introduce an extremum principle
that determines the spontaneous position of the internal wall. Let's
consider an additional external force $F_{w}$ perpendicular to the
wall. In a steady state it is determined by the pressure difference,
$F_{ext}=-A\left(p_{2}-p_{1}\right)$. The force performs work on
the system, $\mkern3mu\mathchar'26\mkern-12mu dW_{ext}=-A\left(p_{2}-p_{1}\right)\,dz_{w}.$
Because the spontaneous direction of the wall is toward the equilibration
of pressures, the system reaches ``equilibrium'' for the position
of the wall $x$, for which the work done by the external force is
negative, 
\begin{equation}
\mkern3mu\mathchar'26\mkern-12mu dW_{ext}<0.\label{eq:minimal work}
\end{equation}
The above phenomenological fact we use to construct the thermodynamic-like
extremum principle. When the upper wall does not move vertically,
the work of the external force is related to the volumetric work in
both subsystems, 
\begin{equation}
\mkern3mu\mathchar'26\mkern-12mu dW_{ext}=-p_{1}dV_{1}-p_{2}dV_{2}.\label{eq:work by pressure}
\end{equation}
This equation, by utilizing the internal energy balance given by Eqs.
(\ref{eq:def dP i through dU dW}) and the nonequilibrium entropy,
is expressed as follows, 

\begin{equation}
\mkern3mu\mathchar'26\mkern-12mu dW_{ext}=dU_{1}-T_{1}^{*}dS_{1}^{*}+dU_{2}-T_{2}^{*}dS_{2}^{*}.\label{eq:elementary work with wall}
\end{equation}
Equivalently, it is given by
\begin{equation}
\mkern3mu\mathchar'26\mkern-12mu dW_{ext}=dU_{12}-\mkern3mu\mathchar'26\mkern-12mu dP_{12},\label{eq:work by U and P whole system}
\end{equation}
where
\begin{equation}
dU_{12}=dU_{1}+dU_{2}\label{eq:dU 1 2}
\end{equation}
and 
\begin{equation}
\mkern3mu\mathchar'26\mkern-12mu dP_{12}=\mkern3mu\mathchar'26\mkern-12mu dP_{1}+\mkern3mu\mathchar'26\mkern-12mu dP_{2}.\label{eq:dP 1 2}
\end{equation}

Eq. (\ref{eq:work by U and P whole system}) puts us in the same position
as in equilibrium thermodynamics: if $\mkern3mu\mathchar'26\mkern-12mu dP_{12}$
has an integrating factor and the corresponding potential, $\mkern3mu\mathchar'26\mkern-12mu dP_{12}=T_{12}^{*}dS_{12}^{*}$,
then the condition of minimal work (\ref{eq:minimal work}) by means
of expression (\ref{eq:work by U and P whole system}) follows the
condition of the minimum of the energy for `adiabatically insulated'
system defined by $S_{12}^{*}=const$ surface. This surface is considered
in the space of $T_{0},V_{1},V_{2},V_{w}$, because we keep the number
of particles , $N_{1},N_{2}$, and other parameters constant. Let's
notice that expressions (\ref{eq:minimal work}-\ref{eq:dP 1 2})
for the case of no shear flow (no heating), $V_{w}=0$, reduces to
the problem of finding the position of the wall that separates two
homogeneous ideal gases. Thus, in equilibrium thermodynamics, $\mkern3mu\mathchar'26\mkern-12mu dP$,
becomes the role the heat differential. The nonequilibrium situation
has been discussed earlier \citep{holyst2022thermodynamics} and it
suggests the minimum principle as follows.

The position of the wall is determined by the minimum of total internal
energy, $U_{12},$ as a function of parameters of states $S_{1}^{*},V_{1},S_{2}^{*},V_{2},$
for the constraint 
\begin{equation}
S_{12}^{*}=S_{1}^{*}+rS_{2}^{*}.\label{eq:total entropy constraint}
\end{equation}
The latter condition denotes steady `adiabatic insulation'. In this
principle there appears parameter $r$. Let's apply the above principle
to verify whether it gives a proper position of the wall and to find
the meaning of $r$ parameter. We apply the above principle with the
constraint of total volume of the system fixed, 
\begin{equation}
V=V_{1}+V_{2}.\label{eq:total volume constraint}
\end{equation}
The total energy is given by
\begin{equation}
U\left(S_{1}^{*},V_{1},S_{2}^{*},V_{2}\right)=U\left(S_{1}^{*},V_{1}\right)+U\left(S_{2}^{*},V_{2}\right),\label{eq:U for both subsystems}
\end{equation}
where $U\left(S_{1}^{*},V_{1}\right)$ follows from the fundamental
relation (\ref{eq:int e fundamental relation}). Minimization of the
above energy with constraints (\ref{eq:total entropy constraint})
and (\ref{eq:total volume constraint}) allows us to use $S_{1}^{*}$
and $V_{1}$ as two independent parameters and minimize the function
$U\left(S_{1}^{*},V_{1}\right)+U\left(\left(S_{12}^{*}-S_{1}^{*}\right)/r,V-V_{1}\right)$.
Its first derivatives lead to the following necessary conditions for
the minimum of the function,
\begin{align}
T_{1}^{*}-\frac{1}{r}T_{2}^{*} & =0,\label{eq:zeroth law}\\
p_{1}-p_{2} & =0.\nonumber 
\end{align}
It proves that the necessary condition for the minimum in the presented
extremum principle leads to equality of pressures. The first of the
above equations gives the interpretation of $r$ parameter. It is
the nonequilibrium temperature ratio. Notice that when $r=1$, the
entropy becomes additive (Eq. (\ref{eq:total entropy constraint}))
and the minimum principle reduces directly to the equilibrium minimum
principle.

In the above reasoning the minimum principle follows from the fact,
that the internal energy balance is determined by $dU_{i}=\mkern3mu\mathchar'26\mkern-12mu dP_{i}-p_{i}dV_{i}$
with homogeneous pressure that is solely determined by the internal
energy and volume, $p_{i}=p_{i}\left(U_{i},V_{i}\right)$. Both assumptions
also hold when transport coefficients depend on parameters of states
\citep{Holyst2023Steady}. In particular, if the viscosity or heat
conductivity depends on density or temperature. For this reason, the
derived minimum energy principle is also valid beyond the regime of
Onsager's linear response theory \citep{Groot_Mazur_Non-equilibrium_thermodynamics}.

It is also worth noting that the inequality (\ref{eq:minimal work})
which determines the direction of spontaneous motion of the wall,
with the use of Eq. (\ref{eq:work by pressure}) and Eq. (\ref{eq:first law total en})
applied to each subsystem is expressed by 
\[
\mkern3mu\mathchar'26\mkern-12mu dW_{ext}=dE_{12}-\mkern3mu\mathchar'26\mkern-12mu dQ_{12}-\mkern3mu\mathchar'26\mkern-12mu dW_{w12},
\]
with energy differential, $dE_{12}=dE_{1}+dE_{2}$, excess heat $\mkern3mu\mathchar'26\mkern-12mu dQ_{12}=\mkern3mu\mathchar'26\mkern-12mu dQ_{1}+\mkern3mu\mathchar'26\mkern-12mu dQ_{2}$
and excess work performed by wall, $\mkern3mu\mathchar'26\mkern-12mu dW_{w12}=\mkern3mu\mathchar'26\mkern-12mu dW_{w1}+\mkern3mu\mathchar'26\mkern-12mu dW_{w2}$.
The above formula suggests an approach to identify the quantity that
is minimized in the system by subtraction from total energy all other
terms (apart from work related to the change of the constraint) that
are present in the energy balance equation (here, excess work performed
by the wall and heat) which describe the ways the system exchanges
energy with the environment. Such a way could lead to a mnemotechnic
rule to identify the physical quantity to generalize the second law
in other situations than considered in this paper (e.g., with the
exchange of particles).

\section{Summary}

In this article, we investigate whether there is a description similar
to equilibrium thermodynamics for out-of-equilibrium steady states.
We consider this problem in the context of the Couette flow, where
there is a mass current (velocity field), energy dissipation, and
a continuous flow of heat. Each feature independently excludes the
theory of equilibrium thermodynamics and its tools.

Since, in general, it is still unclear that such a simple, equilibrium
thermodynamic-like description of nonequilibrium steady states is
possible, we have chosen the most straightforward possible system,
which we believe includes the coupling of mass flow and heat current,
i.e., an ideal gas in shear flow. We show that nonequilibrium entropy
exists, which describes how the system gains energy through excess
heat or dissipation. In addition, the nonequilibrium entropy allows
us to construct the principle of minimum energy, which leads to the
proper state of the system after removing the constraint, which is
the force acting on the wall in the system. It proves the existence
of a description of a system with an out-of-equilibrium heat and mass
flow, which practically takes the form of equilibrium thermodynamics
and reduces to the principle of minimum energy in equilibrium thermodynamics
when the shear flow disappears. 

The thermodynamic-like description introduced in this paper leads
to further questions inspired by equilibrium thermodynamics. Particularly
interesting are the procedures for measuring non-equilibrium state
parameters, effective temperature and entropy, and measuring excess
heat, dissipation, and work the wall performs. It is worth noting
that the obtained result goes along the line of recent experiments
of Yamamoto et al., who extended calorimetry for the measurements
of the excess heat of sheared fluids \citep{yamamoto2023calorimetry}.
Another issue is the possibility of developing the description of
interacting systems, where the fundamental relationship for van der
Waals gas with heat flow can be introduced. It is also interesting
to ask about Maxwell's identities, which in equilibrium thermodynamics
appear very practical. Finally, it is fascinating to check whether
such a description exists for other systems with coupled heat flow
and mass current in the atmosphere and the chemical industry.

\section*{Acknowledgement}

P.J.Z. would like to acknowledge the support of a project that has
received funding from the European Union's Horizon 2020 research and
innovation program under the Marie Sk\l odowska-Curie grant agreement
No. 847413 and was a part of an international co-financed project
founded from the program of the Minister of Science and Higher Education
entitled \textquotedbl PMW\textquotedbl{} in the years 2020--2024;
agreement No. 5005/H2020-MSCA-COFUND/2019/2.

\bibliographystyle{unsrt}

\begin{thebibliography}{10}

\bibitem{ross1988thermodynamics}
John Ross, Katharine~LC Hunt, and Paul~M Hunt.
\newblock Thermodynamics far from equilibrium: Reactions with multiple
  stationary states.
\newblock {\em The Journal of chemical physics}, 88(4):2719--2729, 1988.

\bibitem{hunt1990thermodynamic}
Paul~M Hunt, Katharine~LC Hunt, and John Ross.
\newblock Thermodynamic and stochastic theory for nonequilibrium systems with
  more than one reactive intermediate: Nonautocatalytic or equilibrating
  systems.
\newblock {\em The Journal of chemical physics}, 92(4):2572--2581, 1990.

\bibitem{ross1992thermodynamic}
John Ross, Katharine~LC Hunt, and Paul~M Hunt.
\newblock Thermodynamic and stochastic theory for nonequilibrium systems with
  multiple reactive intermediates: The concept and role of excess work.
\newblock {\em The Journal of chemical physics}, 96(1):618--629, 1992.

\bibitem{Thermodynamics_and_an_Introduction_to_Thermostatistics_2ed_H_Callen}
Herbert~B Callen.
\newblock {\em Thermodynamics and an Introduction to Thermostatistics}.
\newblock John Wiley \& Sons, 2006.

\bibitem{evans1980thermodynamics}
Denis~J Evans and HJM Hanley.
\newblock A thermodynamics of steady homogeneous shear flow.
\newblock {\em Physics Letters A}, 80(2-3):175--177, 1980.

\bibitem{hanley1982thermodynamics}
HJM Hanley and Denis~J Evans.
\newblock A thermodynamics for a system under shear.
\newblock {\em The Journal of Chemical Physics}, 76(6):3225--3232, 1982.

\bibitem{evans1983computer}
Denis~J Evans.
\newblock Computer "experiment" for nonlinear thermodynamics of couette flow.
\newblock {\em The Journal of Chemical Physics}, 78(6):3297--3302, 1983.

\bibitem{evans1986rheology}
Denis~J Evans.
\newblock Rheology and thermodynamics from nonequilibrium molecular dynamics.
\newblock {\em International Journal of Thermophysics}, 7:573--584, 1986.

\bibitem{daivis2003steady}
PJ~Daivis and ML~Matin.
\newblock Steady-state thermodynamics of shearing linear viscoelastic fluids.
\newblock {\em The Journal of chemical physics}, 118(24):11111--11119, 2003.

\bibitem{taniguchi2004steady}
Tooru Taniguchi and Gary~P Morriss.
\newblock Steady shear flow thermodynamics based on a canonical distribution
  approach.
\newblock {\em Physical Review E}, 70(5):056124, 2004.

\bibitem{daivis2008thermodynamic}
Peter~J Daivis.
\newblock Thermodynamic relationships for shearing linear viscoelastic fluids.
\newblock {\em Journal of non-newtonian fluid mechanics}, 152(1-3):120--128,
  2008.

\bibitem{baranyai2018thermodynamic}
Andr{\'a}s Baranyai.
\newblock Thermodynamic integration for the determination of nonequilibrium
  entropy.
\newblock {\em Journal of Molecular Liquids}, 266:472--477, 2018.

\bibitem{nakagawa2019global}
Naoko Nakagawa and Shin-ichi Sasa.
\newblock Global thermodynamics for heat conduction systems.
\newblock {\em Journal of Statistical Physics}, 177(5):825--888, 2019.

\bibitem{nakagawa2022unique}
Naoko Nakagawa and Shin-ichi Sasa.
\newblock Unique extension of the maximum entropy principle to phase
  coexistence in heat conduction.
\newblock {\em Physical Review Research}, 4(3):033155, 2022.

\bibitem{Zhang2021continuous}
Yirui Zhang, Marek Litniewski, Karol Makuch, Pawe\l{}~J. \ifmmode~\dot{Z}\else
  \.{Z}\fi{}uk, Anna Macio\l{}ek, and Robert Ho\l{}yst.
\newblock Continuous nonequilibrium transition driven by heat flow.
\newblock {\em Phys. Rev. E}, 104:024102, Aug 2021.

\bibitem{holyst2022thermodynamics}
Robert Ho\l{}yst, Karol Makuch, Anna Macio\l{}ek, and Pawe\l{}~J
  \ifmmode~\dot{Z}\else \.{Z}\fi{}uk.
\newblock Thermodynamics of stationary states of the ideal gas in a heat flow.
\newblock {\em The Journal of Chemical Physics}, 157(19):194108, 2022.

\bibitem{Holyst2023Steady}
Robert Holyst, Karol Makuch, Anna Maciolek, Konrad Gizynski, and Pawel~J Zuk.
\newblock Steady thermodynamic fundamental relation for the interacting system
  in a heat flow.
\newblock {\em arXiv:2301.12732}, 2023.

\bibitem{oono1998steady}
Yoshitsugu Oono and Marco Paniconi.
\newblock Steady state thermodynamics.
\newblock {\em Progress of Theoretical Physics Supplement}, 130:29--44, 1998.

\bibitem{sasa2006steady}
Shin-ichi Sasa and Hal Tasaki.
\newblock Steady state thermodynamics.
\newblock {\em Journal of statistical physics}, 125(1):125--224, 2006.

\bibitem{seifert2012stochastic}
Udo Seifert.
\newblock Stochastic thermodynamics, fluctuation theorems and molecular
  machines.
\newblock {\em Reports on progress in physics}, 75(12):126001, 2012.

\bibitem{ding2022unified}
Mingnan Ding, Fei Liu, and Xiangjun Xing.
\newblock Unified theory of thermodynamics and stochastic thermodynamics for
  nonlinear langevin systems driven by non-conservative forces.
\newblock {\em Physical Review Research}, 4(4):043125, 2022.

\bibitem{speck2021modeling}
Thomas Speck.
\newblock Modeling of biomolecular machines in non-equilibrium steady states.
\newblock {\em The Journal of Chemical Physics}, 155(23):230901, 2021.

\bibitem{bodenschatz2000recent}
Eberhard Bodenschatz, Werner Pesch, and Guenter Ahlers.
\newblock Recent developments in rayleigh-b{\'e}nard convection.
\newblock {\em Annual review of fluid mechanics}, 32(1):709--778, 2000.

\bibitem{Groot_Mazur_Non-equilibrium_thermodynamics}
Sybren~Ruurds De~Groot and Peter Mazur.
\newblock {\em Non-equilibrium thermodynamics}.
\newblock Courier Corporation, 2013.

\bibitem{Spivak_M_Calculus_on_manifolds}
Michael Spivak.
\newblock {\em Calculus on manifolds: a modern approach to classical theorems
  of advanced calculus}.
\newblock CRC press, 2018.

\bibitem{yamamoto2023calorimetry}
Taro Yamamoto, Yuki Nagae, Tomonari Wakabayashi, Tadashi Kamiyama, and Hal
  Suzuki.
\newblock Calorimetry of phase transitions in liquid crystal 8cb under shear
  flow.
\newblock {\em Soft Matter}, 19(8):1492--1498, 2023.

\end{thebibliography}

\end{document}